\documentclass[12pt,preprint]{aastex}
\def\hone{$\mathrm{H}$\,{\sc i}\,}

\def\mabs{\hbox{$M_{\rm abs}$}\,}

\def\kms{km$\,$s$^{-1}$\,}
\def\magsqarcsec{mag$/$\raisebox{-0.4ex}{\hbox{$\Box^{\prime\prime}$}\,}}
\def\gtsim{~\rlap{\lower -0.5ex\hbox{$>$}}{\lower 0.5ex\hbox{$\sim\,$}}}
\catcode`\"=\active\let"=\"

\def\3{{\ss} }

\def\c12{{1\over 2}}

\def\plusplus{\raise 0.3ex\hbox{${\scriptstyle ++}$}{}}

\shorttitle{Tidal evolution of dSph galaxies}
\shortauthors{Pe\~{n}arrubia,  Navarro \& McConnachie}
\slugcomment{Submitted to ApJ}

\newcommand{\oversim}[2]{\protect{\mbox{\lower0.5ex\vbox{% 
   \baselineskip=0pt\lineskip=0.2ex 
   \ialign{$\mathsurround=0pt #1\hfil##\hfil$\crcr#2\crcr\sim\crcr}}}}}  
\newcommand{\simgreat}{\mbox{$\,\mathrel{\mathpalette\oversim>}\,$}} % >~ sign 
\shorttitle{Tidal stream of NGC4013}
\shortauthors{Mart\'{i}nez-Delgado et al.}

\begin{document}

\title{Discovery of a Giant Stellar Tidal Stream Around the Disk Galaxy
  NGC\,4013}
\author{David Mart\'\i nez-Delgado\altaffilmark{1,2,3}, Michael 
Pohlen\altaffilmark{4,5}, 
R. Jay Gabany\altaffilmark{6}, Steven R. Majewski\altaffilmark{7}, Jorge 
Pe\~narrubia\altaffilmark{8}, Chris Palma\altaffilmark{7,9,10}}

\altaffiltext{1}{Instituto de Astrof\'\i sica de Canarias, La Laguna, Spain.}
\altaffiltext{2}{Max-Planck Institut f\"ur Astronomie, Heidelberg, Germany.}
\altaffiltext{3}{Ram\'on y Cajal Fellow.} 
\altaffiltext{4}{Cardiff University, School of Physics \& Astronomy, Cardiff, 
UK.}
\altaffiltext{5}{Kapteyn Instituut, Rijksuniversiteit Groningen, Groningen, The 
Netherlands.}
\altaffiltext{6}{BlackBird Observatory, Mayhill, New Mexico, USA}
\altaffiltext{7}{Department of Astronomy, University of Virginia, 
Charlottesville, VA 22904-4325, USA.}
\altaffiltext{8}{Department of Physics \& Astronomy, University of Vitoria, 
Canada}
\altaffiltext{9}{Visiting Astronomer, Kitt Peak National Observatory.} 
\altaffiltext{10}{Now at Department of Astronomy, Penn State University, 525 
Davey Lab, University Park, PA 16802}

\begin{abstract}

We report the discovery of a giant,  loop-like stellar
structure around the edge-on spiral galaxy NGC\,4013. This arcing feature
extends 6$\arcmin$ ($\sim$ 26 kpc in projected distance) northeast from the
center and 3$\arcmin$ ($\simeq 12$ kpc) from the disk plane; likely related
features are also apparent on the southwest side of the
disk, extending to 4$\arcmin$ ($\sim$ 17 kpc). The detection of this low
surface-brightness ($\mu_{\rm R}= 27.0^{+0.3}_{-0.2}$ \magsqarcsec) 
structure  is independently confirmed in three separate datasets from three different telescopes. 

Although its true three dimensional geometry is unknown,  the sky- projected 
morphology of this structure displays a match with  the
theoretical predictions for the edge-on, projected view  of a stellar
tidal streams of a dwarf satellite moving in a low inclined ($\simeq
25^\circ$), nearly circular orbit. Using the
recent model of the Monoceros tidal stream in the Milky Way by Pe\~narrubia et
al. as template, we find that the progenitor system may have been a galaxy
with an initial mass
$6\times 10^8 M_\odot$, of which current position and final fate is unknown.
According to this simulation, the tidal stream may be
approximately  $\sim 2.8$ Gyr of age.

Our results demonstrate that NGC\,4013, previously
considered a prototypical isolated disk galaxy in spite of having one of the most prominent \hone warps detected thus far, 
may have in fact suffered a recent minor merger. This discovery highlights that undisturbed disks at high surface brightness levels in the optical but warped in \hone maps may in fact reveal complex signatures of recent accretion events in deep photometric surveys.

\end{abstract}
%
%
%
%
%%%%%%%%%%%%%%%%%%%%%%%%%%%%%%%%%%%%%%%%%%%%%%%%%%%%%%%%%%%%%%%%%%%%%%%5
%%%%%%%%%%%%%%%%%%%%%%%%%%%%%%%%%%%%%%%%%%%%%%%%%%%%%%%%%%%%%%%%%%%%%%%5
\section{INTRODUCTION}
In the last decade, the study of the formation and evolution of the Milky Way
(MW) has been revolutionized by the first generation of wide-field, digital
imaging surveys. The resulting extensive photometric databases have revealed
for the first time the existence of spectacular stellar tidal streams
\citep[e.g., that from the Sagittarius dwarf
  galaxy;][]{Ibata2001a,MD2001,Majewski2003} as well as large stellar
sub-structures in the halo \citep{Newberg2002,RP2004,Juric2005}, interpreted
as the fossils of the hierarchical formation of our Galaxy. The discovery of
the Monoceros tidal stream \citep{Newberg2002,Yanny2003}, located close to the
Galactic plane outside the MW disk \citep[as well as similar structures seen
  around the M31 disk,][]{Ibata2005}, indicates that mergers might play a
relevant role in the formation of the outer regions of spiral disks
\citep{Jorge2006} like that of the MW.  Moreover, inside-out disk formation
from continual accretion of in-falling material is an observed feature in cold 
dark matter ($\Lambda$CDM) simulations of the growth of structures on galaxy-
sized scales
\citep[e.g.,][]{Abadi2003}, and is now also a common feature of galactic
chemical evolution models \citep[e.g.,][]{ALC01, CMR01} attempting to explain
trends in disk chemical abundance patterns.
These various results provide clear evidence that the destruction of satellite
galaxies plays a relevant role not only in the formation of MW-like spiral
galaxies generally, but for their disks as well as their halos.  Furthermore,
these results suggest that the stellar mass assembly of the MW disk, and disks
in general, likely continues actively to the present epoch.

The accretion of satellite galaxies on low-inclination orbits may be an important formation mechanism of galactic disks in the $\Lambda$CDM paradigm \citep[e.g.,][]{Abadi2003}.
The existence of low-inclined
tidal streams in the outer regions of our Galaxy \citep[e.g.,
  the Monoceros or Triangulum/Andromeda tidal
  streams,][]{Yanny2003,Majewski2003,Grillmair2006} supports this prediction, although the external origin of these over-densities is still a matter of debate, mainly due to the severe extinction hindering the exploration of low-latitude areas. Recently, for example,
 Kazantzidis et al. (2008) and Younger et al. (2008) have suggested that, in a $\Lambda$CDM context, the bombardment by dark matter sub-halos of an initially cold disk formed at $z\simeq 1$ would also yield the formation of low-surface brightness substructures ($\mu \simgreat 25$  \magsqarcsec) with filamentary shape that may locally resemble tidal streams in phase-space.
 Clearly, additional observational data (e.g. detailed chemical abundances) obtained along the Monoceros stream are needed to distinguish between both theoretical scenarios.

The search for extragalactic analogies to MW minor mergers is required not
only to (1) show that the MW is not unusual in this respect, but to (2) obtain
externally-viewed ``snapshots'' of different phases of such interactions, (3)
explore the range of possible mass/orbit combinations for such activity, and
(4) estimate the fractional contribution of accreted mass and the mass
spectrum of such events in the life of MW-like systems, an issue that remains
unresolved \citep{Majewski1999}.  In this way, a systematic
survey of tidal streams around other, nearby disk galaxies can provide new
constraints and insights on the hierarchical formation and structure of
MW-like galaxies beyond the previously limited views afforded by our own Galaxy
and M31 \citep{Ibata2007}. A need to build up statistical information on the
number and distribution of tidal streams in galaxies is driven by the
availability of state-of-the-art, high resolution cosmological simulations
that offer, for the the first time, the opportunity to use these observations
to probe the theoretical predictions of galaxy formation in the framework of
the $\Lambda$CDM paradigm \citep[e.g.,][]{Bullock2005}.

Promising galaxies in the hunt for extragalactic tidal streams are those
displaying outstanding asymmetries in optical or \hone images. It has been
long suggested that these perturbations are a result of gravitational
interaction with nearby companions. In some cases, apparently isolated
galaxies exhibit morphologies more commonly associated with interacting
systems. Recently, the prototypical isolated, warped disk galaxy, NGC\,5907,
was found to be surrounded by a spectacular stellar tidal stream
\citep{Shang1998,Martinez-Delgado2008}, but one with no obvious companion
(which may have been completely disrupted).  In the case of both the MW and
M31 there are also obvious disk warps --- both stellar and gaseous --- but no
immediately obvious perturber, though small, nearby, likely tidally disrupting
companion satellites have been suggested as a cause in each case ---
\citep{Sato1986,Ibata1998,Bailin2003,Bailin2004}.
These examples suggest that spirals with warped disks but no large nearby
companions, may have undergone {\em minor} mergers in the last Gyrs.
Under this prepense, we have been observing disk galaxies to faint surface
brightnesses in the search for evidence of minor mergers.  We
\citep{Martinez-Delgado2008} have previously reported our observations of the
debris of a minor merger around the NGC\,5907 disk system.  Here we report 
similar
 work on the edge-on disk, NGC\,4013.
NGC\,4013 is one of the 62 luminous ($M_{B} <-16.9$) members of the Ursa Major
cluster of galaxies, a nearby, late-type dominated, and low mass galaxy
cluster. According to the HYPERLEDA database, the mean heliocentric radial
velocity of NGC\,4013 (836 \kms) places it at a distance of 14.6
Mpc. It is also a relatively isolated system, with the two nearest, slightly
more luminous, cluster neighbours (NGC\,4051 and NGC\,3938) at $\sim170/250$ kpc
projected distance away. Classified as an Sb galaxy with a maximum observed
 rotational
velocity of 195 \kms \citep{Bottema1996}, an extinction corrected total
absolute magnitude of -20.1 \mabs $_{B{\rm band}}$ \citep{Verheijen2001},
 an optical scale-length of 2.8\,kpc (40\arcsec) \citep{vdk1982} and an
 isophote 25 mag arcsec$^{-2}$ radius of 5.2\arcmin \citep{rc3} ,
NGC\,4013 is very similar to the Milky Way.  NGC\,4013 is, moreover, famous for 
its prodigiously
warped \hone disk, with a line of nodes close to parallel with that of
the line of sight and with one of the largest warp angles observed ($\sim
25\deg$).  On one side the warp extends out to about 8\,kpc (2$\arcmin$) from 
the nominal plane \citep{Bottema1987,Bottema1995,Bottema1996}. 
The box/peanut shaped bulge of NGC\,4013 with its 'X'-like morphology
indicates the presence of a bar being observed edge-on \citep{Patsis2006}.

Here we report the optical detection of a faint, low-surface brightness,
loop-like structure that appears to be part of a giant, low-inclination
stellar tidal stream of a disrupted dwarf satellite looking very similar to
the expected form of the Monoceros tidal stream. This looping structure
suggests the likelihood of there having been a previous interaction of
NGC\,4013 with a low-mass companion. Given the fact that NGC\,4013's very
prominently warped \hone layer matches in shape and orientation those of the
apparent tidal loop makes this galaxy system a very compelling example of the
likely link between disk warps and mergers, and an interesting case study for
models of warp formation.
%
%
%%%%%%%%%%%%%%%%%%%%%%%%%%%%%%%%%%%%%%%%%%%%%%%%%%%%%%%%%%%%%%%%%%%%%%%5
%%%%%%%%%%%%%%%%%%%%%%%%%%%%%%%%%%%%%%%%%%%%%%%%%%%%%%%%%%%%%%%%%%%%%%%5
\section{OBSERVATIONS AND DATA REDUCTION}
\label{data}

We observed NGC 4013 with three different telescopes and instruments. In each
case the same structures appeared, albeit with different degrees of
clarity.

\subsection{KPNO 0.9-m Telescope}

NGC\,4013 was initially observed with the Kitt Peak National Observatory
(KPNO) (now WIYN) 0.9-m, f/8 telescope as part of a pilot survey of low
surface brightness features (like stellar tidal streams) around warped, nearby
disk galaxies by C.P. and S.R.M.  The initial sample included NGC\,3044,
NGC\,3079, NGC\,3432 and NGC\,4013, all edge-on systems of large angular size
with \hone warps but no nearby, massive companions (though, in the cases of
NGC\,3432 and NGC\,3079, some associated or nearby low surface brightness or
dwarf satellites).  These disk systems resemble NGC\,5907, which at the time
of this pilot survey had already shown to have a tidal loop by
\cite{Shang1998}, and, in the case of NGC\,3044, a minor merger was already
hypothesized \citep{Lee1997}.

The observations were made with the Mosaic camera during several nights (UT
2000-03-31 to 2000-04-04) using the BATC 9 filter --- essentially a narrow
version of the Cousins $R$ band --- used by \cite{Shang1998} in their imaging
of NGC 5907; the filter was kindly loaned by Rogier Windhorst. The Mosaic
camera is made of eight 2k $\times$ 2k individual chips and provides a field
of view (FOV) of 59$\arcmin$ at a pixel scale of 0.43\arcsec pixel$^{-1}$. The
conditions for this KPNO run were photometric, with a typical seeing of $\sim$
1.3 $\arcsec$. Using standard IRAF procedures for the over-scan/bias correction
and flat-fielding, we combined the final image (see upper panel of
Fig.\ref{KPNOINTimage}) of NGC\,4013 from the sum of four 1200\,sec exposures,
eight 1800\,sec exposures, and one 900\,sec exposure.

A very faint arc is clearly identified in this image, situated at $\sim$
6$\arcmin$ northeast of the center of the galaxy in the region of the extreme
of northern plume of the gaseous warp \citep[][see
  Sec.~\ref{discussion}]{Bottema1996}. However, as may be seen, the achievable
flat-fielding and surface brightness limit was not adequate to make definitive
conclusions about the nature of this structure.

\subsection{Isaac Newton Telescope}

To confirm and better trace the extent of this structure, follow-up
observations were obtained with the 2.5m Isaac Newton Telescope (INT) at La
Palma using the Wide Field Camera (WFC) at f/3.29. This instrument holds four
$4096\arcmin\!  \times\!2048\arcmin$ pixels EEV CCDs with a pixel size of
$0.332\arcsec$, providing a total field of about
$35\arcmin\!\times\!35\arcmin$. Images were acquired during two service nights
(UT 2003-03-27 and 2003-04-03) using the Sloan $r'$ (\#214) and Harris $B$
(\#191) filters. From the first night we have two 600\,s $r'$ band exposures
and from the second night three additional 600\,s $r'$ band images plus four
1350\,s $B$ band exposures.

The central CCD (chip 4) was large enough (FOV
$22.8\arcmin\!\times\!11.4\arcmin$) to include NGC\,4013 and its warp, so we
reduced only chip 4 in isolation, using standard IRAF procedures for the
over-scan/bias correction, flat-fielding, and combining of the individual
images. The background on the final $B$ band image turned out to be flat and
we obtained a sky value of 22.22 B-\magsqarcsec from ellipse fits, using the
fixed outer disk geometry, beyond the galaxy as described in
\cite{Pohlen2006}. The resulting 3-sigma uncertainty on the sky value is below
0.04\%.  However, the individual $R$ band exposures suffered from scattered
light causing large-scale variations in the background of the order of 2\%.
After careful masking of all sources, which were replaced by the mean of a
linear interpolation along lines and columns, we used the IRAF {\it imsurfit}
routine on a median filtered ($\sim 30\arcsec$ window) version of the masked
and interpolated image to determine the background on each individual one.
After subtracting this second order fit from the images they were combined to
the final $r'$ band version reducing the large-scale variations in the
background to 0.2\%. The two images used in this study have an equivalent
exposure time of 50\,min for the $r'$ band and 90\,min for the $B$ band.  Due
to the unstable weather conditions (high humidity, sometimes cirrus) during
the two nights we did not use the observed standard stars, but obtained our
photometric calibration from the $r'$ and $g'$ band Sloan Digital Sky Survey
(SDSS) \citep{York2000} by means of aperture photometry.  After geometrically
mapping the two INT and two SDSS images to the same scale and center and
applying an identically-sized, but conservative (i.e.~large) mask on the
disturbing, bright star close to the center of NGC\,4013, we obtained fluxes
in six concentric apertures (40\arcsec-240\arcsec diameter) on all four
images. Following \cite{Smith2002} we converted the magnitudes measured on the
SDSS images to Cousins $R$ and Johnson $B$, and obtained the photometric
zero-points for our INT images using a linear fit without a color term.  The
uncertainty is of the order of 0.01 mag. We get similar results by using
aperture photometry of six isolated, bright, but unsaturated stars. All the
following magnitudes are given in $R$ and $B$, correcting for Galactic
extinction, \citep[$A_B=0.072$, $A_R=0.044$;][]{Schlegel1998}, but not for
inclination.

The final image of NGC 4013 is shown in the lower panel of Figure
\ref{KPNOINTimage}. Although the background is still an issue we can now more 
confidently trace and measure the low surface brightness arc. It is important
to remark that all quoted surface brightness measurements in this paper (see
Sec. 3) come
from the data set obtained at this telescope, the only one where we
have calibrated, deep enough  photometric data.

%
%%%%%%%%%%%%%%%%%%%%%%%%%%%%%%%%%%%%%%%%%%%%%%%%%%%%%%%%%%%
\subsection{Black Bird Remote Observatory Telescope}

The evidence of a possible tidal stream around NGC 4013 from our previous data
obtained with the KPNO telescope and INT  motivates
us to select this galaxy as a priority target in our pilot survey of stellar
tidal streams in nearby spiral galaxies using small telescopes, of which
searching strategy was successful proven with the detection of very faint
parts of the NGC 5907 tidal stream (the commissioning target of this
survey; see Mart\'inez-Delgado et al. 2008). 

With this purpose, we obtained very deep images of NGC4013 with the f/8.3
Ritchey-Chretien .5-meter telescope of the Black Bird Remote Observatory
(BBRO) situated in the Sacramento Mountains (New Mexico, USA).  The data was
obtained during several dark sky observing runs between UT 2006-11-06 and UT
2006-12-28. We used a Santa Barbara Instrument Group (SBIG)
STL-11000 CCD camera, which yields a large field of view (27.7$\arcmin$ x
18.2$\arcmin$) and a plate scale of 0.45\arcsec pixel$^{-1}$. The data
consists of multiple deep exposures through non-infrared clear luminance
 (CL; 3500$<\lambda <$ 8500), red, green and blue filters from the SBIG Custom
Scientific filters set. Table 1 gives a summary of the images collected for
this project. Column 5 refers to the total exposure time of the co-added
images for each filter obtained with each run. This yields a total exposure
time per filter of 5400s (red), 3240s (green),  6480s (blue) and 39600s (CL). 

A master dark and bias frame was created by combining 10 dark sub-exposures
each produced at the same exposure length and camera temperature settings used
for the luminance and the filtered exposures. A master flat was produced by
combining 10 separate sky flat exposures for each filter. The science data
sub-exposures were reduced using standard procedures for bias correction and
flat fielding. The data captured through the CL filter was median
combined to produce a master CL-filter data set. The red, green and blue
filtered exposures were separately combined (using the median technique) to
produce individual red, green and blue master image that represented the
total exposure time through each color filter.  The resulting data sets for each
color filter were subsequently summed to produce a ``synthetic'' luminance
image
 that represented the total exposure time accumulated through all of the
color filters, that was subsequently summed combined with
the  CL-filter image . The
resulting final image thus represented the
sum of all available CCD exposures collected in this project, with an accumulated exposure
time of 15.2 hours (including 11 hours through a clear luminance filter).

The dynamic range of the final image was very
large, because of the presence of a very bright, galactic disk surrounded by
very diffuse tidal stream. To optimize the contrast of the faint structures detected
around NGC 4013, we used an  histogram equalization of the image by means
of a non-linear transfer
function\footnote{see
  http://http://homepages.inf.ed.ac.uk/rbf/HIPR2/histeq.htm and references
  within.}. This well-known image processing tecnique employs a monotonic,
 non-linear mapping which re-assigns the
 intensity values of pixels in the input image such that the output image
 contains an uniform distribution of intensities. This yields an effective
 method to suppress the bright regions of the image, intensifying
 the fainter parts of the stream. In addition, to increase the signal-to-noise of the faint structures surrounding NGC4013,
image noise effects were reduced by the application of a Gaussian
filter \citep{Davies1990}.An illustrative example of the effect of these
processes is given in Fig. 2.

The final image of NGC 4013 was obtained using an  iterative process  that involves
several passes of a  S-shaped histogram operator to the total image and 
the application of a three-pixel diameter low-pass Gaussian  filter. This approach
optimized contrast and detail throughout the tonal range of the image and permitted
the faint structure surrounding NGC 4013 to be rendered distinguishable in the
final image. The resulting image is shown in Figure~\ref{BBRO1} . 
We have added the labels A through G
to identify some photometric feature we discuss in the following Section.

 As it is discussed in Mart\'inez-Delgado et al. (2008), the search strategy
of this survey was designed to obtain the position and morphology
of very faint structures around the spiral galaxy, with the aim of providing a
guide for pencil-beam, follow-up photometric surveys in state-of-art
telescopes. Because of this approach is mainly based on the
use of a very broad-band, uncalibrated luminance filter,  the
BBRO telescope data is only used in this paper  to discuss the overall
morphology of the possible stream. From our previous experience (e.g. NGC
5907: Mart\'inez-Delgado et al. 2008; Messier 94: Trujillo et al. 2008, in preparation), we estimate that the final image should
include diffuse light structures as faint as $\mu_{\rm R}\sim 29$
\magsqarcsec .

\section{THE TIDAL STREAM(S) OF NGC\,4013}
\label{stream}
\begin{figure}
\includegraphics[width=10cm,angle=270]{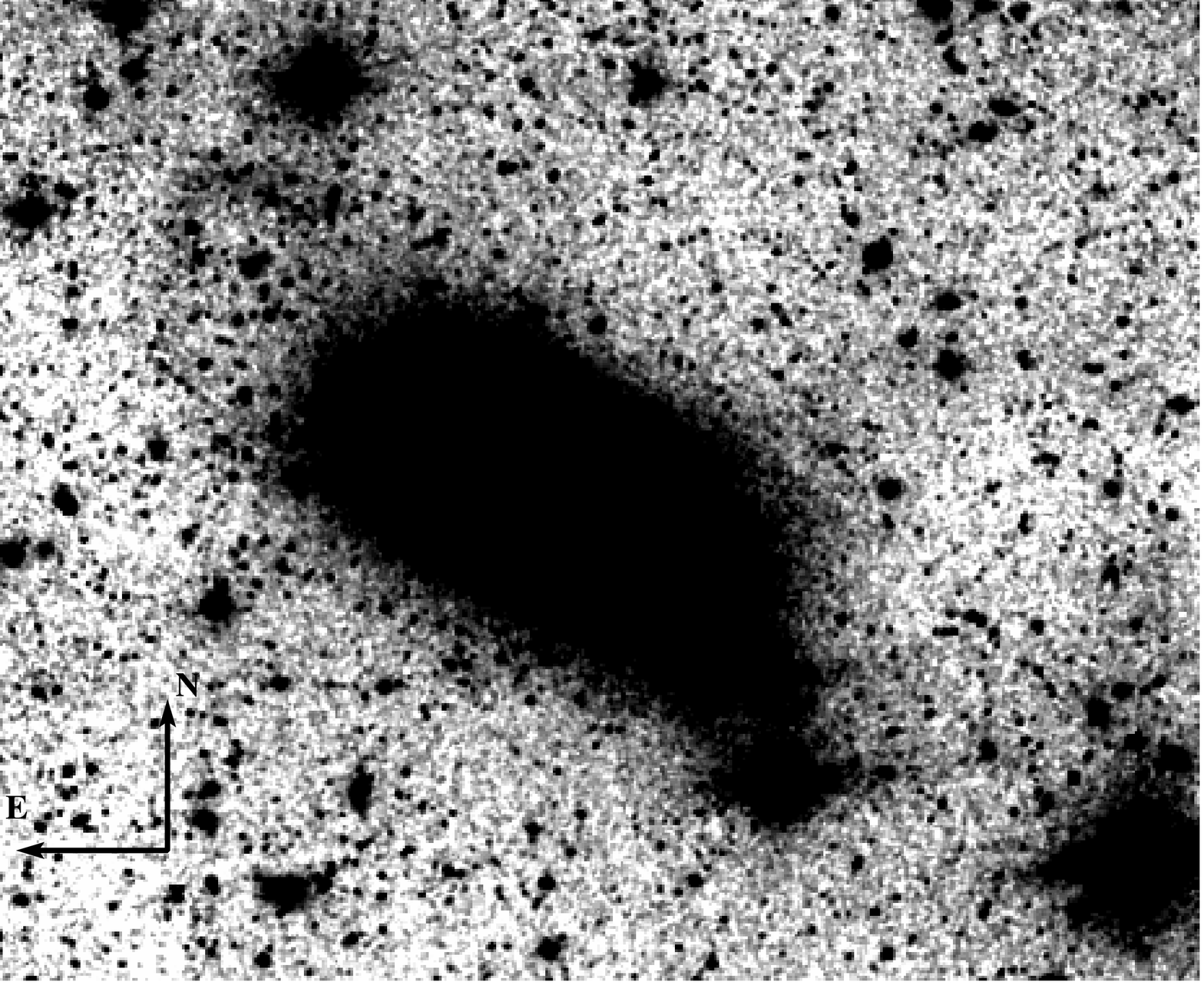}
\includegraphics[width=10.8cm,angle=270]{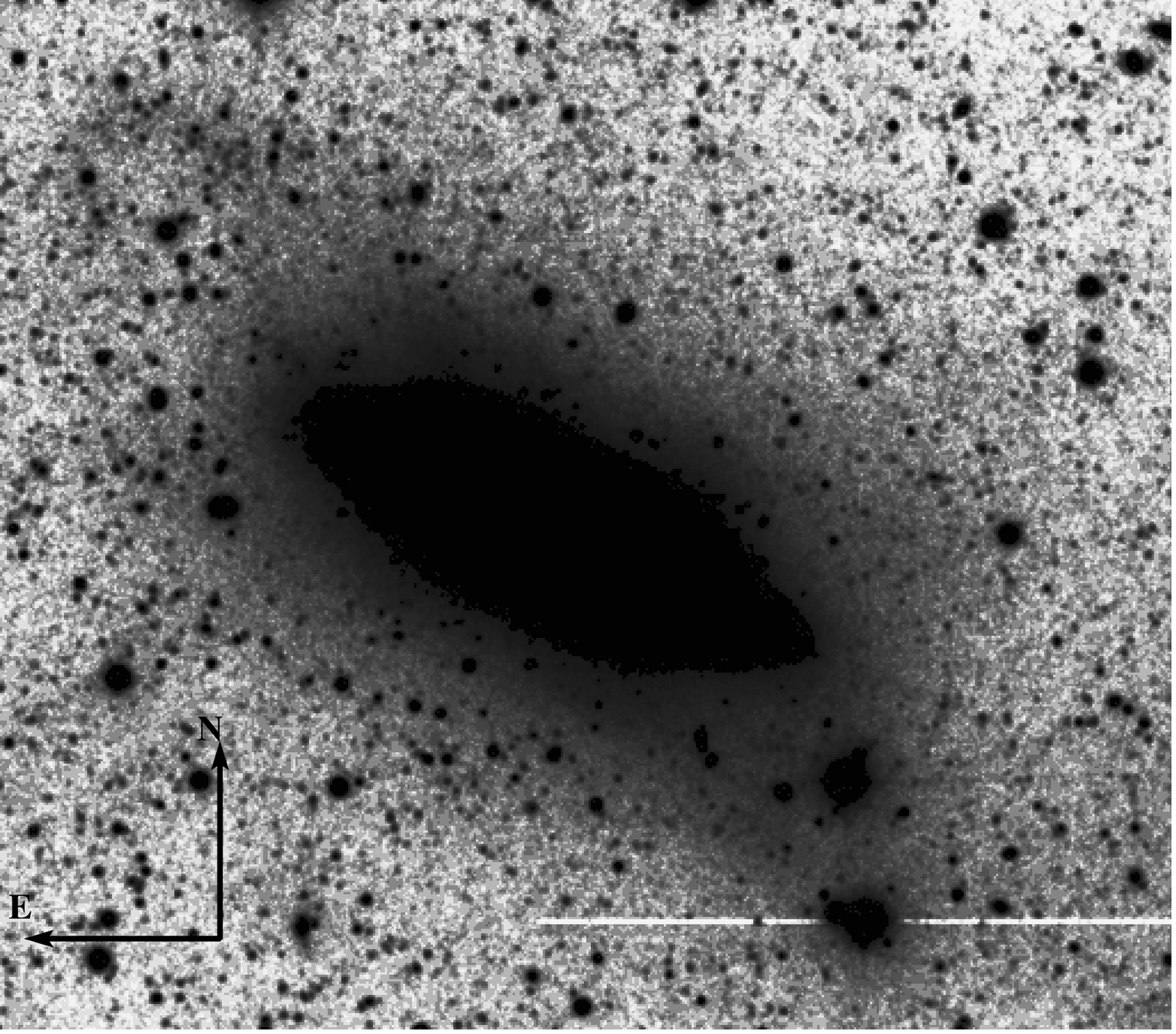}
\caption{NGC\,4013: Smoothed and enhanced versions of the KPNO 0.9-m image
  {\it(upper panel)} and the INT $B$ band image {\it(lower panel)} highlighting
  the low-surface brightness features. The arrows of lengths $2\arcmin$ in the
  lower left corner give the size and orientation. Both images are displayed
in logarithm scale.}
\label{KPNOINTimage}
\end{figure}

In addition to the well known peanut-shaped bulge \citep[][feature A in our
  Fig.~\ref{BBRO1}]{vdk1982} with a full vertical extent of about 2\,kpc 
(30$\arcsec$), a contour plot of the INT image (Fig.~\ref{INTcontour}) reveals
a huge, at least 13\,kpc ($>3\arcmin$) sized box-shaped outer ``stellar halo''
(feature B in Fig.~\ref{BBRO1} and also obvious in the images in
Fig.\ref{KPNOINTimage}). Furthermore, to the East, a striking horseshoe-like
structure (feature C in Fig. \ref{BBRO2}) is visible in the BBRO image,
starting at the northeast corner of the box-shaped halo and reaching out to at
least 26\,kpc (6\arcmin) from the center. This feature, that corresponds to
the arc detected in the KPNO and INT images (see Sec.~\ref{data} and
Fig.\ref{KPNOINTimage}), is now evident as loop-like. The obvious hole in the
light distribution clearly shows that this feature is not simply a stellar
warp, but an actual ``ring-like" structure seen obliquely, and of the type
produced by tidal streams (see Sec.~4). The loop apparently enters the galaxy
again at the southeast corner of the box (labeled F in Fig.~\ref{BBRO1}).

The BBRO and INT images also show a pair of {\it wings} jutting to the west and
southwest (features D and E) and straddling this side of the NGC 4013 disk,
{\it squaring off} the most diffuse galaxy light distribution at large radii.
 In the diffuse light visible in the northwest edge of the galaxy (feature E),
there is some evidence for a shorter, coherent loop feature visible in a
high-contrasted version of the BBRO image (Fig.~\ref{BBRO2}). This is
consistent with the presence of a second arm of debris (as expected in
theoretical simulations, see Sec.~4), which is probably related to the
northeast loop (feature C). However, deeper images are necessary to confirm
this hypothesis. The more plume-like feature visible on the southwest side
(feature D in Fig.~\ref{BBRO1}) extends $\sim$ 17 kpc (4$\arcmin$) out. 
 Unfortunately, the clarity of this feature (also barely visible in the
INT image in Fig.~\ref{KPNOINTimage}) is somewhat hindered by the presence of
two bright stars.

Together, these wings, and the obvious loop to the northeast could be seen as
parts of one continuous {\it ribbon-bow} or  {\it pretzel-like} structure
surrounding the disk but with arms that cross each other at the galaxy center(at least in
projection). Evidently, these particular features of the extended light
distribution are the result of a single, coherent tidal disruption event,
which is creating a low-inclination stellar stream around the disk of
NGC\,4013.  An example of how such a tidal structure might be formed is shown
in a tidal disruption model (see Figure \ref{fig:FoS} and discussion in
Section \ref{discussion}).  It is worth noting that we do not observe any
bright spot that could be identified with a remaining, intact dwarf galaxy
core among any of the low-surface brightness features. Therefore, our images
do not provide any evidence on the final fate of the progenitor galaxy, which
could be hidden inside the disk or could be completely disrupted by now.

The surface brightness of the described structures is very close to the
remaining inhomogeneities in the background of our CCD INT images, specially
on the $r'$-band image, which makes it extremely difficult to determine
accurate surface brightnesses. Using a set of small $\approx4\arcsec\!x\!4\arcsec$ boxes, strategically placed along the stream by avoiding stars and any residual
obvious brightness condensation, we estimated the mean surface brightness of
the stream on the east side (feature C) to be about $\mu_{\rm B} =
28.6^{+0.6}_{-0.4}$ and $\mu_{\rm R}=27.0^{+0.3}_{-0.2}$. The large errors
are due to the uncertainty of $\sim0.1$\% in the local sky determination (a
set of similar small boxes placed outside, but close to the stream) and are of
the same order as the brightness variation along the stream. This yields a
$({\rm B}-{\rm R})$ color of $1.6^{+0.6}_{-0.4}$, which is a typically red
value found for S0 galaxies \citep[e.g.][]{Barway2005}, and would be
consistent with the redder dwarfs of the Local Group \citep{Mateo1998}.
Although the color of the loop is about 0.6\,mag redder than the outer
(radially/vertically) parts of the disk/halo of NGC\,4013 and only close to
the mid-plane of the galaxy (where the dust lane certainly reddens the
intrinsic colors), we find similar colors as for the stream within the attached
large error. Therefore, deeper broad-band observations are  needed to
undoubtedly conclude whether  its stellar population is not composed of the
same mix of stars found in the outer region of NGC\,4013's disk, an additional evidence that this horseshoe-like feature did not originate in the disk itself.

The width of this giant stream is variable along its path, but this may be due
to a projection effect. A rough estimate for the full-width-half-maximum (FWHM) of the stream
$1.0\!\pm\!0.3 $ kpc ($\sim 14$\arcsec) is taken at the most distant part on the
northeast side of feature C in Fig.~\ref{BBRO1}.

Figure \ref{BBROpHI} shows the position of the possible tidal stream with
respect to the prominent \hone gas warp \citep{Bottema1987,Bottema1996}.  This
comparison clearly shows that there is no \hone gas associated with the
detected giant stellar feature \citep[such as is the case in NGC 
 3310;][]{Wehner2005} on the northeast side (feature C). On this side, the
projected path of the stream shows a very similar inclination to the warped
gas disk, but is clearly outside and almost enclosing (in projection) the
measured \hone gas

Interestingly, our high-contrasted version of the BBRO image
(Fig.~\ref{BBRO2}) reveals the trace of a spike-like feature in the northeast
edge of the galaxy disk (feature G), with a position and orientation with
respect to the galactic plane that is consistent with the stellar counterpart
of the prodigious gaseous warp. It is clear, however, that (at our surface
brightness limit) this stellar component warp is significant less extended
than the \hone-component displayed in Fig.\ref{BBROpHI}.  On the southeast
side the extended optical feature D (see Fig.~\ref{BBRO1}) seems to be
associated to the southern \hone gas plume of the warp, although on this side
we are severely hampered by the two bright foreground stars.  However, the
clear separation of the \hone gas and the optical light on the other side of
the galaxy let us reject the direct association of these structures as an
extended, low luminosity stellar counterpart of the warped \hone disk of
NGC\,4013.
%
%
%%%%%%%%%%%%%%%%%%%%%%%%%%%%%%%%%%%%%%%%%%%%%%%%%%%%%%%%%%%%%%%%%%%%%%%5
%%%%%%%%%%%%%%%%%%%%%%%%%%%%%%%%%%%%%%%%%%%%%%%%%%%%%%%%%%%%%%%%%%%%%%%5

\section{DISCUSSION}
\label{discussion}

Our deep images of NGC\,4013 show a complex stellar halo, including a
coherent, stream-like structure inclined $\sim\!25\deg$ with respect to the
galaxy disk. We have independently confirmed the presence of these low
surface brightness features with three data sets from three different
telescopes. In addition, if one knows what to look for, the loop is also
barely visible on the Digitized Sky Survey 2 (DSS2) blue plate and on the archival
$4.5\mu$ Spitzer Infrared Telescope Facility image. The existence of the loop
structure is therefore without question. The discovered structures are
evidently debris material produced by the tidal disruption of a low-mass
galaxy companion to NGC 4013: Such features are expected from tidal disruption
of a companion galaxy \citep{Johnston2001} and similar features are also
observed in NGC\,5907 \citep{Shang1998,Martinez-Delgado2008}, NGC\,3310
\citep{Wehner2005}, and a handful of other external galaxies \citep[summarized
  in][Table 1]{Martinez-Delgado2008}.

The sky-projected geometry of this structure is in fact  similar to
the edge-on, external perspective of a tidal stream that results from the tidal disruption of 
a single dwarf galaxy satellite moving on a low-inclined orbit  predicted
by different theoretical simulations (e.g. the Monoceros tidal stream: Pe\~narrubia et al. 2005;Canis Major over-density: Martin et al. 2005, their
Fig. 14). For comparison purpose,  we will use here as  template the
N-body model constructed by Pe\~narrubia et al. (2005)  to reproduce the
geometry and kinematics of the Monoceros tidal stream. This model simulates the accretion of a low-mass
satellite (6 $\times$10$^{8}$ M$_{\sun}$) on a nearly circular 
(e=0.10$\pm$0.05), low inclined ($i\!\sim\!25^{\circ}$) orbit in the last
3\,Gyr and provides a remarkable good match to the projected, edge-on view of
the stream wrapping NGC 4013 (see Fig.~\ref{fig:FoS}). Considering that NGC\,4013 has a
smaller disk scale-length, $h_R\!=\!2.8$ kpc, and a lower maximum circular
velocity, $V_{c,{\rm max}}\!=\!195$ \kms \citep{Bottema1996} than the Milky
Way, we can apply dynamical considerations to estimate the stream age from the N-body model, so that ${\rm Age}/{\rm
  Age}_{Mon}= h_R/h_{R,MW}\times  V_{c,{\rm max}}/V_{c,{\rm max},MW}$. For the Milky
Way we have that $h_R=3.5$ kpc and $V_{c,{\rm max}}=220$ km/s, thus ${\rm
  Age}/{\rm Age}_{Mon}\approx 0.9$. Since some pieces of the Monoceros stream
are ${\rm Age}_{Mon}=3$ Gyr old, we estimate that the stream surrounding
NGC\,4013 may be as old as $\simeq 2.8$ Gyr. However, although the model provides a reasonable qualitative resemblance to the morphology of NGC 4013, it
cannot fully explain the observed structure. For example, while the observed tidal structure in Fig. 3  is clearly  more prominent on its Eastern side of the galaxy, the model shows loops of roughly equal brightness on both sides (as well as various other features that not detected in our deep image). Unfortunately, with no kinematic data and only the projected geometry of the stream as constraint, the orbit and mass of the progenitor system may be degenerated, and in that respect the estimates presented here must be considered illustrative only. 

Although the progenitor system of the Monoceros and NGC 4013 tidal streams
might have shared similar orbits and
initial masses, our estimates reveal that the surface brightness of the latter
is approximately $\sim$2--3 magnitudes
higher than the value reported for this low galactic latitude tidal
debris candidate ($\mu_{\rm B}\sim 30.0 $\magsqarcsec: J. de Jong, private communication).
A possible explanation might be
that the progenitor of the NGC\,4013 stream was more luminous than that of the
Monoceros stream. Since both models have the same initial mass, this hypothesis would
imply that the mass-to-light ratio of the NGC\,4013 stream progenitor was
considerably lower than that of the Monoceros stream.  
However, the N-body models used by Pe\~narrubia et al. (2005) assume that the stellar and dark matter components of the progenitor system have the same spatial distribution (i.e. mass-follows-light models). 
In a $\Lambda$CDM cosmogony, stars only populate the
inner-most regions of dwarf galaxies (Strigari et al. 2007, Pe\~narrubia et al. 2008a) and can only be tidally
stripped after most of the dark matter halo beyond the luminous radius has
been lost (Pe\~narrubia et al. 2008b), which introduces an additional
parameter: the spatial segregation of the stellar component with respect to the
surrounding dark matter halo of the progenitor galaxy, which is fundamental to
determine the survival time of dwarf galaxies and the properties of their
associated tidal streams.  For example, the more
deeply embedded within the halo a stellar component is, the thinner, brighter
and colder will be its associated tidal stream. Therefore, for $\Lambda$CDM-
motivated
models, the halo size, the stellar spatial segregation and the total
luminosity of a dwarf galaxy are all parameters that influence the resulting
surface brightness of an associated tidal stream but which,  unfortunately, cannot be constrained in
absence of additional data (e.g. kinematics).

Whereas the detection of the stream around NGC\,4013 cannot shed light on the
nature of the Milky Way stellar over-densities (including on the controversial
origin of the Monoceros stream; see Sec. 1), it does suggest that the accretion of satellite galaxies on low-inclination, low-eccentric orbit may be relatively common during the hierarchical formation of spiral galaxies.
In fact,
their occurrence in the known sample of stellar tidal streams in the Local
Volume \citep[$D\!<\!15$\,Mpc; see][]{Martinez-Delgado2008,Pohlen2004} is
$\gtsim 20\%$. In this context, 
 the presence of giant disks around spiral galaxies that extend
out to several scale-lengths of the inner disk (e.g.~the extended disk found by
\cite{Ibata2005} around M31 is detectable out to a projected radius of 80\,kpc
with the present instrumentation) might also be signature of the accretion of
massive satellite galaxies moving on low-inclination, low-eccentricity orbits
\citep{Jorge2006}, adding support to the scenario that the disks of spiral
galaxies may grow inside out as a result of this type of accretion events.

The detection of galactic warps in spiral galaxies that present signatures of
recent accretion events is quickly increasing with the availability of
wide-field, deep photometric surveys. Although the origin of galactic warps is
still unknown, it was early proposed that the gravitational perturbations
induced by satellite galaxies may cause the formation of warps
(Burke 1986, although see Hunter \& Toomre 1969 for counter-arguments).
Since those early studies, several alternative scenarios have been proposed
in the literature. For example, it has been shown that the interaction of
a stellar disk with the surrounding dark matter halo may also induces the
formation of warps \citep{Sparke1988,Binney1998}, as well as bending 
instabilities \citep{Revaz2004}, intergalactic accretion flows onto the disk 
\citep{Lopez2002} and cosmic
in-fall \citep{Shen2006}. 

A crucial fact that has been often invoked to dismiss the tidal origin of
galactic warps was the existence of warped spiral galaxies in apparent
isolation \citep{Sancisi1976}. NGC\,4013, with its prominent, and rather
symmetrical \hone warp, along with a really unassailable indication of a past
merger event, provides an important counter-example to the above argument;
indeed, NGC\,4013 may well be the Rosetta Stone for warp theories, since
morphologically the warp and the merger debris seem so closely aligned.

With NGC\,5907 and NGC\,4013 we have now two examples of apparently isolated
galaxies with significantly warped gaseous disks, which at the same time show
evidence for an ongoing tidal disruption of a dwarf companion. The connection
between the warp of the MW and its satellite galaxies has been already
explored in some recent studies. For example, \cite{Bailin2003} and
\cite{Weinberg2006} study the effects of the Sagittarius dwarf and the
Magellanic Clouds on the MW disk, respectively. Interestingly, the
existence of warped galaxies that have suffered a recent accretion of
satellite galaxies on low-inclination, low-eccentricity orbits (like the MW and 
NGC\,4013) suggests that these kind of accretion events may induce the
formation of warps more efficiently than the perturbations from satellite
galaxies moving on highly eccentric, nearly-polar orbits. This is a
possibility worth to be further investigated.

With the growing numbers of examples showing a connection between warped disks
and evidence of mergers, even for disk systems with no obvious companions, we
may be close to resolving the mystery of warps, and in a way be consistent with
prevailing $\Lambda$CDM models of galaxy formation.
\acknowledgements
We would like to thank Roelof Bottema for making the \hone contour map
available to us. D.~M-D acknowledge funding from the Spanish Ministry of
Education (Ramon y Cajal program contract and research project AYA
2004-08260-C03-02). C.P. and S.R.M. thank Rogier Windhorst for the loan of the
BATC 9 filter for the KPNO 0.9-m observations.  Part of this work was
supported by a Marie Curie Intra-European Fellowship within the 6th European
Community Framework Programme. C.P. and S.R.M. thank Rogier Windhorst for the
loan of the BATC 9 filter for the KPNO 0.9-m observations. S.R.M. acknowledges
support from National Science Foundation grants AST 97-02521 and AST 03-07851,
a Cottrell Scholar Award from the Research Corporation, NASA/JPL contract
1228235, the David and Lucile Packard Foundation, and the generous support of
The F. H. Levinson Fund of the Peninsula Community Foundation.  JP thanks
Julio F. Navarro for finantial support. Funding for the creation and distribution
of the SDSS Archive has been provided by the Alfred P. Sloan Foundation, the
Participating Institutions, the National Aeronautics and Space Administration,
the National Science Foundation, the U.S. Department of Energy, the Japanese
Monbukagakusho, and the Max Planck Society. The SDSS Web site is {\ttfamily
  http://www.sdss.org/}. This research has made use of the Online Digitized
Sky Surveys (DSS1 \& 2) server at the ESO/ST-ECF Archive produced by the Space
Telescope Science Institute through its Guide Star Survey group.

%

%%%%%%%%%%%%%%%%%%%%%%%%%%%%%%%%%%%%%%%%%%%%%%%%%%%%%%%%%%%%%%%%%%%
%%%%%%%%%%%%%%%%%%%%%%%%%%%%%%%%%%%%%%%%%%%%%%%%%%%%%%%%%%%%%%%%%%%

\clearpage
   
\begin{deluxetable}{lclccc} 
\tablecaption{JOURNAL OF OBSERVATIONS (BLACK BIRD OBSERVATORY)}  
\tablewidth{15cm} 
\tablehead{\colhead{Date} & & \colhead{Filter} & N$_{f}$ &  Exp.Time (s) &
  \colhead{Total Exp. Time (s)}} 
\startdata 
2006 November 20 & &CL & 8 & 800 & 7200\\ 
2006 December 17   & & CL & 7 & 1800 & 12600\\ 
2006 December 24   & & Red & 2 & 900 & 1800\\ 
2006 December 24   &  &Green & 1 & 540 & 540\\ 
2006 December 24   &  &Blue & 1 & 1080 & 1080\\ 
2006 December 27   &   &CL & 8 & 1800 & 14400\\ 
2006 December 28   & &Red & 4 &900 & 3600\\ 
2006 December 28  & & Green & 5 & 540 & 2700 \\  
2006 December 28  & & Blue & 5 &1080& 5400\\ 
\hline 
\enddata 
%\tablecomments{ } 
\label{obslog} 
\end{deluxetable} 

\clearpage

\begin{figure}
\centering
\includegraphics[width=.55\textwidth,angle=0]{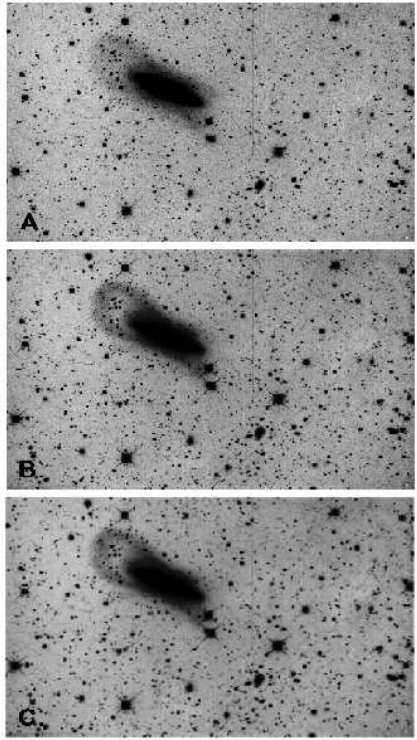}
\caption{ An illustrative example of the effect of histogram equalization and
  Gaussian filtering techniques in the resulting final image obtained from the sum of all
  the exposures taken with the BBRO: A) the image with linear-stretching (and thus non-proccesed) ;
  B) the image  after applying an histogram equalization with a non-linear function; and C)  the effect of a
Gaussian filter of 5 pixels of radius in the summed image displayed in panel A.
The blurrer aspect of this image with respect to the final version displayed in
Fig. 3 is due the larger filter widthness used  here to enhancent the effect
of the filter with illustrative purposes. The field of view of
these images is $ 28 \arcmin  \times 18 \arcmin$. }
%\label{INTcontour}
\end{figure}

\begin{figure}
\includegraphics[width=15cm,angle=0]{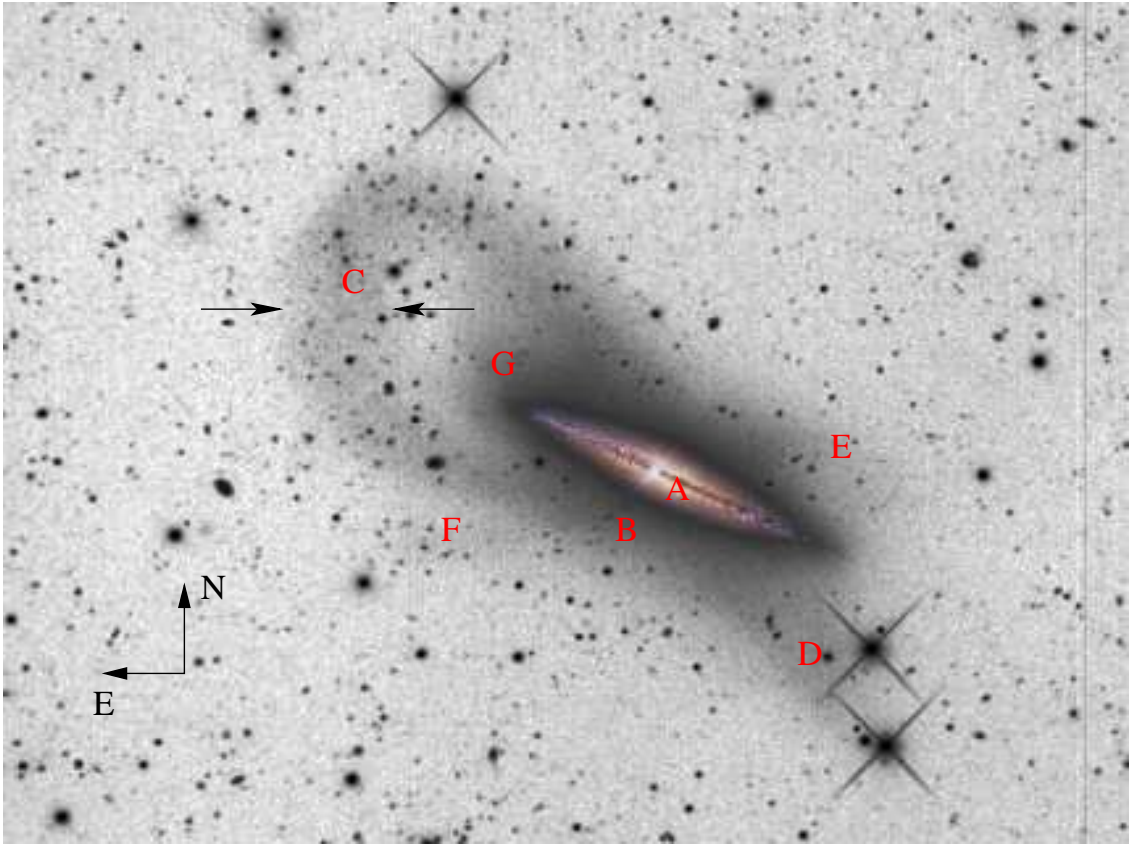}
\caption{Image of NGC4013 obtained with the BBR0 20-inch telescope. The total
  exposure time of the original image was 13.7 hours (including 11 hours
in clear-luminance filter)  and was noise-filtered
  by applying a Gaussian filter.  The image has dimensions of $\sim
  16.8\arcmin \times 12.8\arcmin$, which, at the distance of NGC 4013 is $\sim$
  71 $\times$ 54 kpc. Identified photometric features labeled A-G are
  discussed in Sec.\ref{stream}. For illustrative purpose, a color image of
  the NGC 4013 obtained with the same telescope has been superimposed on the
  saturated disk region of the galaxy. }
\label{BBRO1}
\end{figure}

\begin{figure}
\hspace*{-1cm}
\includegraphics[width=12cm,angle=270]{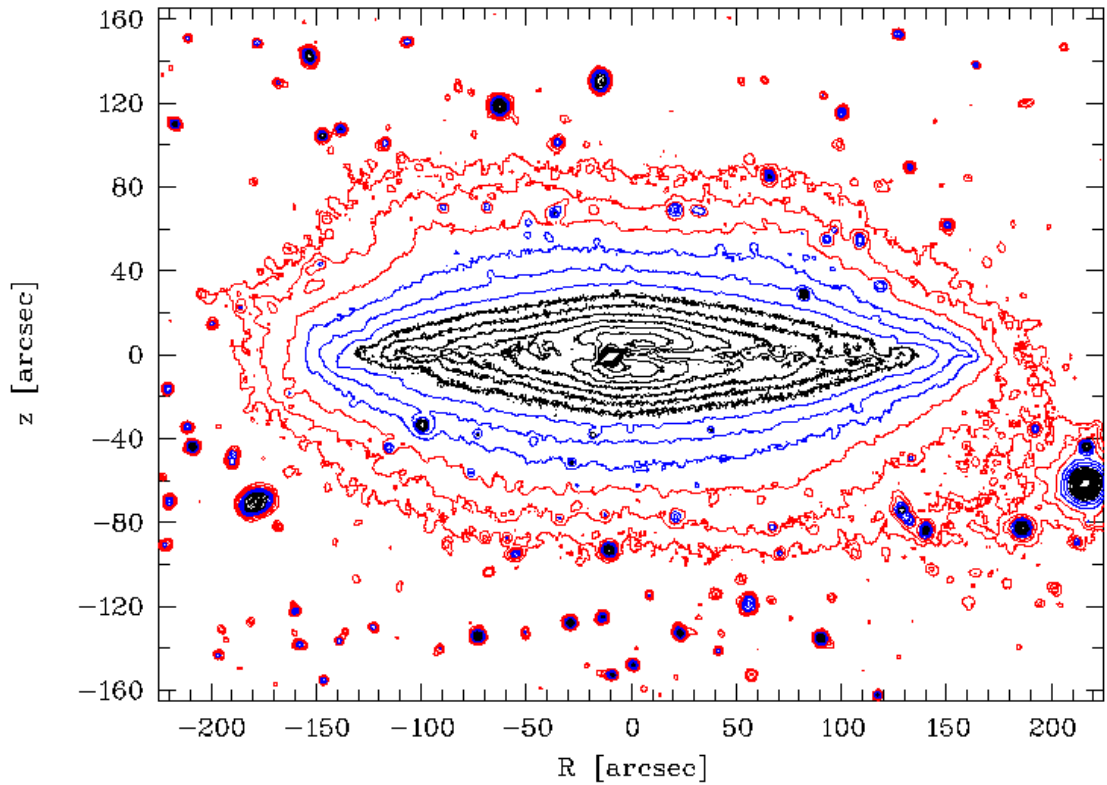}
\caption{Isaac Newton Telescope $B$ band contour map of NGC\,4013 highlighting
  the box-shaped outer halo. Contours are at levels 17.0-27.0 B-mag
  arcsec$^{-2}$ in equidistant steps of 0.5\,mag. Towards the outer parts we
  smoothed the contours using three increasing levels of median filtering
  (indicated by the three different colors).}
\label{INTcontour}
\end{figure}

\begin{figure}
\includegraphics[width=10cm,angle=270]{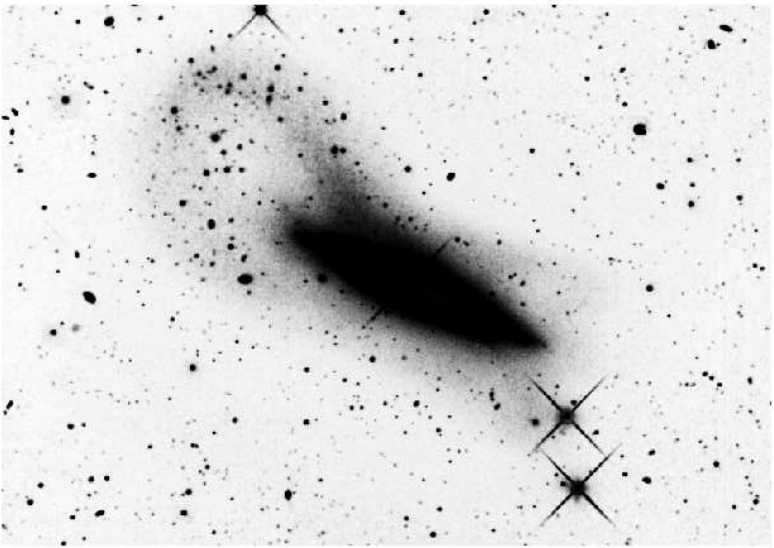}
\caption{ A high- stretched version of the luminance clear-filter image of NGC\,4013
  obtained with the BBRO telescope. This shows evidence of a possible second
  arm of debris within the diffuse light wind to the northwest edge of the
  NGC4013 disk (feature E in Fig.\ref{BBRO1}). In addition, the spike-like
  feature visible in the northeast edge of the disk could be related to the
  stellar component of the prodigious gaseous warp of the galaxy.  North is up
  and East to the left.}
\label{BBRO2}
\end{figure}

\begin{figure}[]
%\vspace{20cm}
%\special{psfile=f6.eps angle=0 hscale=100 vscale=100 hoffset=-100 voffset=-100}
%\plotone{ngc4013_f6.ps}
%\vspace*{-1.0cm}
\hspace*{-3.5cm}
\includegraphics[width=16.5cm]{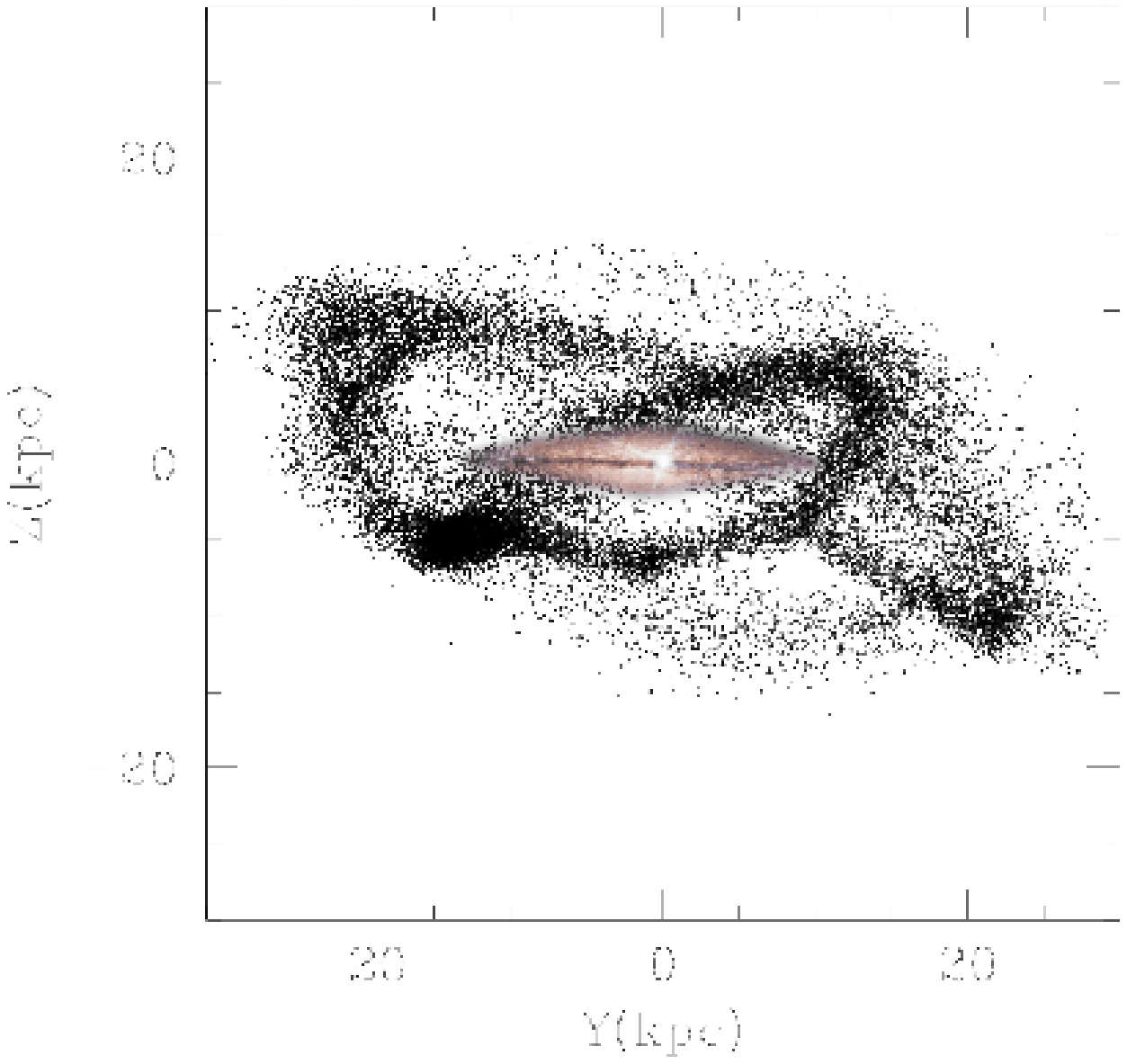}
\caption{X-Y galaxy plane projection of the prograde model {\it pro1} of the
  Monoceros tidal stream by \cite{Jorge2005}. For comparison purposes, a
  color image of the NGC\,4013 disk (see Fig.\ref{BBRO1}) has been scaled
  (assuming a distance of 14.6\,Mpc) and superimposed to the distribution of
  debris. The comparison of this model snapshot to the structures detected in
  NGC 4013 provides considerable support for the hypothesis that they consist
  of different pieces from a single tidal stream. The NGC\,4013 disk (and the
  residual light of its box-shaped outer halo showed in Fig.\ref{BBRO2}, not
  simulated here) almost overlaps the full path of the stream, that emerges as
  individual loop-like structures of debris (or winds) at large galactocentric
  radius (features C and E in Fig.\ref{BBRO1} and \ref{BBRO2}}
\label{fig:FoS}
\end{figure}

\begin{figure}
\hspace*{-3.5cm}
\includegraphics[width=16.5cm,angle=270]{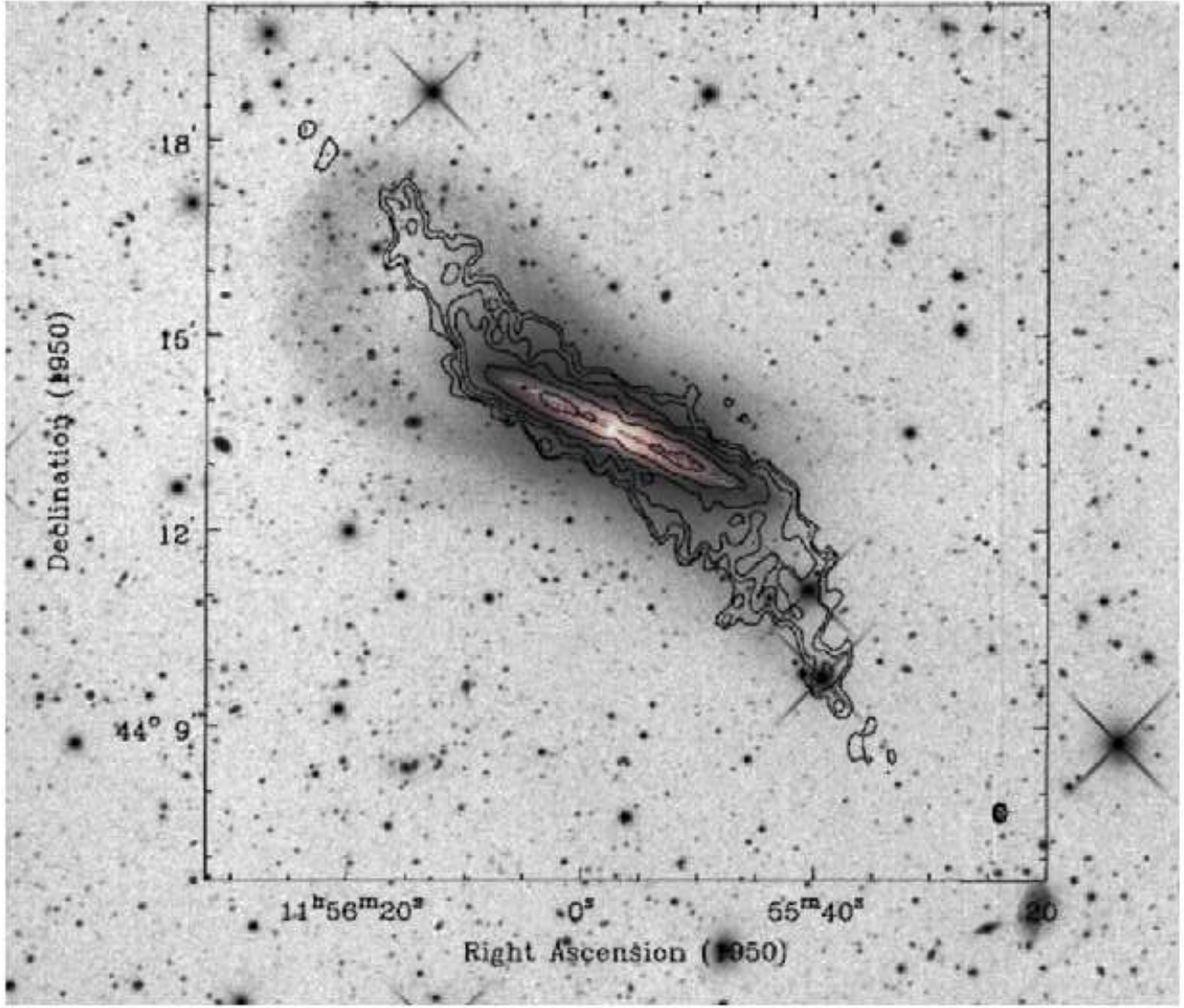}
\caption{Overlay of the \hone contours from \cite{Bottema1996} (his Fig.1) on
  our BBO image of NGC\,4013. The lowest \hone contour level shown is about
  $1\!\times\!10^{20}$ H-atoms cm$^{-1}$. The beam is given in the lower right 
corner
  of the \hone\ insert. The scale of this image is $\sim\!4.2$ kpc/arcmin.}
\label{BBROpHI}
\end{figure}

\end{document}